\documentstyle[12pt]{article}

\title{\bf Observable Dirac-type singularities in Berry's phase
and the monopole }

\vspace{60mm}

\author{\bf Rajendra~Bhandari}

\date{ }

\begin{document}

\maketitle
\vspace{40mm}
\begin{center}
\begin{tabular}{ll}
            & Raman Research Institute, \\
            & Bangalore 560 080, India. \\
            & email: bhandari@rri.res.in\\
\end{tabular}
\end{center}
\vspace{40mm}
-----------------------------------------------------------------------\\
Submitted to Physical Review Letters. Revised version of 21 November 2001.

\newpage
\begin{center}
{\bf Abstract}
\end{center}
\vspace{1mm}

A three-dimensional generalization of the sign-change
($\pi$ phase shift) rule for adiabatic cycles of spin-1/2 or 
two-state  wavefunctions encircling a degeneracy in the parameter 
space of the hamiltonian yields a Dirac-type singularity wherein 
any closed circuit of the adiabatic cycle in which the 
degeneracy is ``looped", results in an observable $\pm 2\pi$ 
phase shift. It is concluded that an interferometer loop taken 
around a magnetic monopole of strength n/2 yields an observable 
$\pm 2n\pi$ phase shift, $n$ being an integer.

\newpage
\section{Introduction:}
\label{sec-introd}

In 1959, Aharonov and Bohm \cite{ab} made the important observation
that a topological phase factor $e^{i\phi}$ picked up 
by an electron moving in a closed circuit around a magnetic field, 
introduced by Dirac in his 1931 paper on the monopole \cite{dirac}, 
is a measurable physical effect, $\phi$ being proportional to the 
magnetic flux through the circuit. Wu and Yang \cite{wuyang} 
conjectured that only the phase factor $e^{i\phi}$ and not 
the phase $\phi$ itself is measurable, implying that $\phi$ is 
defined only modulo $2\pi$. 
More recently, in another well known work, Berry \cite{berry1} discovered 
a topological phase ({\it geometric phase}) in the evolution of a 
quantum system under the action of a cyclic, adiabatic hamiltonian 
and related it to the Aharonov-Bohm phase. Berry's phase, as well as 
the nonadiabatic geometric phases discovered by 
Pancharatnam \cite{panch1}, Aharonov and Anandan \cite{aa}, 
Samuel and Bhandari \cite{jsrb} etc. have implicitly or explicitly been
defined as modulo $2\pi$ quantities.

A different perspective on the geometric phase has been brought out 
in a series of experimental and theoretical contributions by the 
present author \cite{nonint,jumps,rbdirac,4pism,iwbs,rbreview,rbaps}. 
This work uses interference of polarized light, exploits the 
mathematical isomorphism between polarization of light and the 
two-state quantum system and shows through experiments that
(1) a continuously measured geometric phase shift is unbounded,
as opposed to modulo 2$\pi$,
and can be nonintegrable on the parameter space of the
experiment \cite{nonint},
(2) a geometric phase shift as defined by the Pancharatnam 
criterion \cite{panch1} can have discontinuous jumps and can 
change sign for small variation in the parameters
near singular points (or lines, surfaces) in the parameter space
where the two interfering states become orthogonal 
\cite{jumps,4pism,iwbs} and (3) a circuit in parameter space enclosing
several such singularities results in a measurable phase shift 
$\int d\phi$ equal
to 2$\pi$ times the algebraic sum of the strengths of the
singularities \cite{rbdirac,iwbs}, hence the term
``Dirac singularities" \cite{dirac}.
These results have a bearing on the question of observability of
$2n\pi$ phase shifts and add a new dimension to the $4\pi$
spinor symmetry problem \cite{4pism}.

The above results were obtained in the context of nonadiabatic
quantum evolution \cite{aa}, using Pancharatnam's definition of 
phase difference between different states \cite{panch1}. 
In a geometric description, the phase jumps are easily understood 
in terms of the ``shortest geodesic rule" for closing open paths 
in the state space \cite{jumps,remark}.
In this paper, with the help of a model problem
very close to the original adiabatic setting in which
Berry's geometric phase was arrived at \cite{berry1,berry3}, 
we show that an observable Dirac-type singularity is also 
inherent in the adiabatic geometric phase which is like the 
phase acquired by a charged particle going around a loop in 
the field of a magnetic monopole \cite{berry1}.

\section{The Model Problem:}

A beam of quantum mechanical spin-1/2 particles
(electrons or neutrons), in a spin state $\mid{\psi}_i>$,
enters a ring-shaped configuration of paths at some point i
(Fig.1) such that it has a choice of two paths 1 or 2
through one or the other identical halves of the ring,
under the action of spin-hamiltonians $H_1$ or $H_2$ and exits
at the diametrically opposite point f in state $\mid{\psi}_{f1}>$
or $\mid{\psi}_{f2}>$ . The latter, for $m=1,2$, are given by,
\begin{equation}
{\mid{\psi}_{fm}>} ~= ~ U_m{\mid{\psi}_i> ~=~
T exp[(-i/\hbar)\int H_m(t)dt]~~\mid{\psi}_i>}.~~~~~~~~~ \label{eq:evol}
\end{equation}
\noindent Here $H_m (t) ~= ~ \mu(\vec{\sigma}. {{\vec{B}}_m}(t))$, 
$\mu$ is the gyromagnetic ratio and $\vec{\sigma}$ is vector
of Pauli matrices.
The magnetic field ${\vec{B}}_m (t)$ seen by the particle at 
time t consists of two parts:
(1) a constant field with components $B_1$ and $B_z$
in the x and z directions respectively and (2) a rotating field
in the x-y plane with constant magnitude $B_0$ and
frequency $\omega$, $\omega$ being such that $\omega$T=$\pm \pi$;
T being the traversal time for paths 1 and 2. The
rotation has the opposite sense in paths 1 and 2.
Thus in equation (1), $B_{mx}=(B_1+B_0cos ~\omega t)$,
$B_{mz}=B_{z}$, $B_{1y}=(B_0 sin ~\omega t)$ and $B_{2y}=(-B_0 sin 
~\omega t)$. For $B_z=0$, the problem reduces to
that studied by Geller \cite{lyanda}
who proposed a mesoscopic physics experiment to detect an abrupt
$\pi$ phase jump resulting from an adiabatic circuit
encircling a degeneracy in the parameter space
of the hamiltonian; the
``rotating magnetic field" term in their problem arising from
spin-orbit interaction of the electrons. The current J at the other end
of the ring is proportional to ${\mid<\psi\mid \psi> \mid}^2$, where
$\mid \psi>=(\mid{\psi}_{f1}>+\mid{\psi}_{f2}>) $.
The term of interest in the expression for J is the
modulus $c$ and phase $\alpha$ of the
interference term $2<{\psi}_{f2}\mid{\psi}_{f1}>$,
the Pancharatnam phase difference between $\mid{\psi}_{f1}>$
and $\mid{\psi}_{f2}>$. In general, both $c$ and $\alpha$
vary as the parameters
in $H_1$ and $H_2$ are varied. Experimentally, $\alpha$ can be determined
by introducing a variable, state-independent phase difference between the
two halves of the ring till J is maximum.
This is routinely done in neutron interferometry by rotating
a phase shifter  in the path of one or
both the beams, e.g. in  recent geometric phase experiments
\cite{wr2}.

Let us define dimensionless variables $b_1=B_1/B_0$, $b_z=B_z/B_0$,
$\beta=(\mu B_0)/(\hbar \omega)$ and $\gamma=(\mu B_z)/(\hbar \omega)$.
Consider an experiment in which the variables $b_1$ and $b_z$ 
are varied along 
a path such as ABCDA, EFGHE or SPQRS in Fig. 2 by appropriate 
variation of the fields $B_1$ and $B_z$, while the phase difference 
$\alpha$ is being continuously monitored. When $B_z=0$ and $B_1=\pm B_0$,
i.e. the points  $S_1$ and $S_2$ in Fig. 2, the adiabatic cycle 
passes through the point of degeneracy (the point O in Fig. 1)
and these are the singular points. Crossing of any one of these 
points by the adiabatic cycle results in a phase jump of magnitude 
$\pi$. However, in this case there is inevitable departure from 
adiabatic evolution near the singularities. Exactly at the 
singularity, the two final states $\mid{\psi}_{f1}>$
and $\mid{\psi}_{f2}>$ are orthogonal. This is the case 
studied in ref. \cite{lyanda}.

When $B_z \neq 0$, a closed nonadiabatic
solution for the evolution of the wavefunctions along
paths 1 and 2 under the above hamiltonian does not exist.
However, for the central result of this paper, we do not need 
the nonadiabatic solution. When the  circuit 
in the parameter space stays a finite distance away from the 
singularity, it is  
possible to choose $\omega$ small enough so that the 
evolution is adiabatic over the entire cycle all along the 
circuit. For simple circuits of the kind considered (Fig. 2), 
this is true if $\gamma \gg 1$. In the limit $\gamma \rightarrow \infty $, 
the two final states $\mid{\psi}_{f1}>$ and $\mid{\psi}_{f2}>$
are the same and the dynamical phase in the two paths 
exactly cancels, making the problem equivalent to that 
of a single full cycle along the ring with the dynamical phase 
subtracted  \cite{berry1}.
The phase $\alpha$, a purely geometric phase, is then given 
by the solid angle subtended by the adiabatic cycle at the 
degeneracy. If one considers the evolution of the projection 
of the cycle on a unit sphere centered at the degeneracy,
it is easy to see that any closed circuit 
of the adiabatic cycle of the type shown in Fig. 2, i.e. 
a circuit that ``loops" the degeneracy, sweeps an area equal 
to the entire sphere, i.e a solid angle $\pm 4\pi$, the 
sign depending upon the sense of traversal of the circuit.
For circuits like ABCDA and EFGHE which pass close to 
the singularity, the sharp variation of the area of the 
projected cycle on the unit sphere when the periphery of 
the adiabatic cycle is closest to the degeneracy (the point 
O in Fig. 1) can be intuitively seen. The change 
in the sense of this variation with the sign of $B_z$ can 
also be visualized easily. The resulting phase shifts 
are then defined unambiguously and do not have a ``modulo $2\pi$" 
ambiguity at any stage.

We have carried out numerical simulations of the 
actual evolution of the wavefunctions along paths 
1 and 2 under the action of the hamiltonian $H_m$ 
and computed the variation of $\alpha$, i.e. the quantity 
$\int d\alpha$, along a few circuits 
in the space of parameters $b_1$ and $b_z$, e.g. 
the circuits ABCDA, EFGHE and SPQRS in Fig. 2.
Along each  of the four segments of a rectangular circuit, 
100 equally spaced points are chosen and 
for each value of  $b_1$ and $b_z$, the unitary time evolution
operator for a time interval $\delta t =\pi/(20000\omega)$ 
is computed at each of 20000 equi-spaced points along paths
1 and 2. The adiabaticity parameter 
$\beta$ has been chosen so that $\gamma = (b_z \beta) = 20$  
for all the circuits, i.e. for the circuit ABCDA, $\beta = 2000$,
for EFGHE, $\beta = 200$ and for SPQRS, $\beta = 20$. 
The products of the 20000 U matrices along each path 
are then computed to yield $U_m$, which, multiplied 
with the initial wavefunction $\mid{\psi}_i>$ (taken to be 
an eigenstate of $H_m(0)$), yields the final states $\mid{\psi}_{f1}>$
and $\mid{\psi}_{f2}>$. The phase difference $\alpha$ is then 
computed according to the expressions
given above. In the adiabatic limit the dynamical phase in
the two paths is exactly compensated and $\alpha$ reflects
the geometric phase difference. 

Fig. 3 shows the results of the computation. The computed
variation of $\alpha$ shows all the expected features 
mentioned above. The net phase change equal to $-2\pi$, 
the sharp phase jumps with the expected relative sign 
at the two points where the cycle passes close to the 
degeneracy (circuits ABCDA and EFGHE), the increase 
in sharpness of the jumps for a closer approach to the 
degeneracy and the smooth variation of the phase when the 
cycle always remains far from the degeneracy (SPQRS) are 
all clearly seen. We have verified that the sign of the 
phase shifts reverse with the sense of traversal of the 
circuits. For several circuits that do not enclose the 
singularity, the computed phase change is found to be zero. 
It is also found that as $\gamma$ increases, 
the final states $\mid{\psi}_{f1}>$ and $\mid{\psi}_{f2}>$ 
are closer to each other  and that all the 
three curves shown in Fig. 3 approach those for the  variation 
of the solid angle subtended by the cycle at the degeneracy. 

The above results suggest a useful 3-dimensional generalization 
of the sign-change rule. If the adiabatic cycle in the parameter
space is taken through a closed circuit such that it encircles the
degeneracy in the process, the wavefunction acquires 
a $+2\pi$ or $-2\pi$ phase change depending upon the sense of the circuit. 
If it does not encircle the degeneracy, the phase change is zero.
The generalization suggested by Berry \cite{berry1},
based on Stone's result \cite{stone}, refers to a segment
like PQ (with P $\rightarrow \infty$, Q $\rightarrow -\infty$) in
Fig. 2. The
present one is a more complete statement and may find 
uses in molecular problems. We also note that for the special
case of the present problem ($B_z=0$) studied by Geller \cite{lyanda} 
our results agree with theirs. We have shown that for $B_z\neq 0$, 
for the same direction of motion of the loop,
their $\pi$ phase jump has a negative sign for $B_z > 0$ and 
a positive sign for $B_z < 0$.
This is nontrivial. One could for example set up a contraption 
such that a $+\pi$ phase shift accumulated in an electronic 
register activates a switch that triggers an explosive that 
kills a cat ... while a $-\pi$ phase shift does not.

\section{The Monopole:}

Motivated by the above results, consider the following gedanken
experiment with a Dirac monopole. Take a current loop
in the x-y plane, divided into two halves similar to those
in Fig. 1, which now represents the real space with space coordinates 
x, y, z replacing the magnetic field components $B_x, B_y, B_z$. 
Let the radius of the loop be $B_0$, the x-coordinate of its centre
be $B_1$ and its height above the x-y plane be $B_z$.
A current of charged spinless quantum  particles
enters at point i, has a choice of two paths 1 and 2 and
is recombined at the point f where the phase difference $\alpha$
between the two
complex transmission amplitudes $a_1$ and $a_2$ is measured
by an interference experiment. Let a magnetic monopole of strength 
$-1/2$ be
located at the point O (Fig. 1) and the current loop
be transported along the path ABCDA (or EFGHE or SPQRS) as shown, while
$\alpha$ is being continuously monitored and
the orientation of the loop in space kept unaltered.
It is easy to convince oneself that the
variation in $\alpha$ due to the changing magnetic flux through
the loop (Aharonov-Bohm effect), given by half the solid
angle subtended by the loop at O, would be similar to that
shown in Fig. 3, implying  Dirac-type singularities in the
$b_1-b_z$ plane at the points $S_1$ and $S_2$ (Fig. 2). 
If, as usually assumed, the magnetic monopole has a 
string attached to it, there would be an additional $+2\pi$ 
phase jump when the loop encircles the string. This would 
be infinitely sharp hence unobservable if the string is 
infinitely thin but would 
be observable, leading to a net zero phase change for the 
circuit, if the string had a finite thickness.

The above results have been obtained using spin-1/2 wavefunctions
evolving under an $\vec{S}. \vec{B}$ hamiltonian, $\vec{S}$ 
being the spin vector.
However, it is an exact mathematical result \cite{berry1} that 
the phase change of a wavefunction with spin component n/2 along 
the field direction, evolving under the same 
hamiltonian, is exactly n times that for a wavefunction with spin 
component 1/2. We conclude 
therefore that the phase change measured by the loop taken around 
a monopole of strength $n$/2, with an infinitely thin string, 
equals $\pm 2n\pi$, which is thus intrinsic to the problem and 
cannot be truncated to its modulo $2\pi$ value which is zero.

It may also be useful to note that for particles with
spin  quantum number $n$/2, with $n$ $>$ 1 and for problems involving 
more than two quantum states,
the monopole picture of the geometric phase would not be
valid for arbitrary hamiltonians.  One could, however,
expect measurable phase jumps equal to $\pm n\pi$ in general. 

\section{The proposed neutron experiment:}

A set of two counterrotating magnetic fields in the
two arms of a neutron interferometer in  planes normal
to the beams can be set up with the technique used in the
polarimetric experiment of Bitter
and Dubbers \cite{bitter}, along with a uniform but variable
magnetic field $B_1$ normal to the plane of the interferometer and
a magnetic field $B_z$ along each of the beams. The phase shifts as a
function of $B_1$ and $B_z$ can be measured with the
technique used in ref. \cite{wr2}. The measurement
of the current J, i.e. the flux of neutrons in the
recombined beam as a function of $B_1$ and $B_z$ is of course
straightforward. It is also important to note that the phase difference
between  $\mid{\psi}_{f1}>$ and $\mid{\psi}_{f2}>$ at each point
in the parameter space is determined
modulo $2\pi$. For large $\beta$, therefore, it is very
sensitive to small fractional errors in the dynamical phase
in path 1 or 2. The basic topological effect can, however,
be seen for smaller values of $\beta$ by choice of a circuit
that does not pass too close to the singularity.

\newpage

\section*{}{\Large {\bf Figure Captions}}\\


{\bf Figure 1:}
For a given set of parameters $b_1$ and $b_z$, the
propagating particle sees a spin-hamiltonian corresponding
to a magnetic field rotating in one sense, represented by
the semicircle i1f, if it goes through the right half of
the ring and to one rotating in the opposite sense, i.e. i2f
if it goes through the left half. A change in
$B_1$ or $B_z$, or both, moves the circle i1f2i, resulting
in general in a change in the final states
$\mid{\psi}_{f1}>$ and $\mid{\psi}_{f2}>$,
in the phase difference $\alpha$ between them and consequently 
in the current J. \\

{\bf Figure 2:}
Circuits in the parameter space $b_1,b_z$ (not to scale) chosen 
for computing phase shifts shown in
Fig. 3. $S_1$ and $S_2$ are singular points where
$\mid{\psi}_{f1}>$ and $\mid{\psi}_{f2}>$ become orthogonal and
the phase between the two beams becomes undefined.
For adiabatic evolution, a counterclockwise (clockwise) 
circuit enclosing the singularity $S_1$
gives a phase change $- 2\pi$ ($+2\pi$).\\

{\bf Figure 3:}
The solid, the dotted and the dot-dash lines
show the change in phase difference $\alpha$
between the two beams in final states
$\mid{\psi}_{f1}>$ and $\mid{\psi}_{f2}>$,
as the parameters $b_1$ and $b_z$
are varied along the circuits ABCDA, EFGHE and SPQRS (Fig. 2) respectively.
Each segment of the circuit is divided into 100 equi-spaced points
for computation. Note the total phase shift equal to $-2\pi$
and the sharp variation of phase when
the circuit passes close to the singularity, i.e. near the points 50
and 250 for circuits ABCDA and EFGHE. \\

\end{document}